# Magnomechanically induced transparency in the ferrimagnetic bridge crystal of atom opto-magnomechanical system

WENTING DIAO[1,†], XI WANG[2,†], KE DI[2,3], YU LIU[2], ANYU CHENG[2], CHUNXIAO CAI[1], WENHAI YANG[1] AND JIAJIA DU[2*]

[1]National Key Laboratory of Science and Technology on Space Microwave, China Academy of Space Technology Xi'an, 710100, China
[2]Chongqing University of Post and Telecommunications, Chongqing, 400065, China
[3] State Key Laboratory of Quantum Optics and Quantum Optics Devices, Shanxi University, Taiyuan, 030006, China
[†]These authors contributed equally.
*dujj@cqupt.edu.cn

**Abstract:** We investigate the absorption and transmission properties of a weak probe field in an atom opto-magnomechanics system. The system comprises an assembly of two-level atoms and a magnon mode within a ferrimagnetic crystal, which directly interacts with an optical cavity mode through the crystal's deformation displacement. We observe optomechanically induced transparency (OMIT) via radiation pressure and a magnomechanically induced transparency (MMIT) due to the nonlinear magnon-phonon interaction. In addition, due to the coupling of the atom to the detected and signal light, the system's width transparency window is divided into two narrow windows. Additionally, we demonstrate that the group delay is contingent upon the tunability of the magnon-phonon coupling strength. Our solution possesses significant in the field of quantum precision measurement.

**1. Introduction**

Optomechanical and cavity magnomechanics systems have been approved both theoretically demonstrate an analogy to three-level atomic electromagnetically generated transparency [1-4]. This interaction is fundamentally nonlinear at the quantum level [5]. Optomechanics explores the interaction between a laser-driven optical cavity and a vibrating end mirror. The light response of this system also undergoes changes [6-7], which can result in a variety of effects, including electromagnetic induced transparency (EIT) and slow and stopped light [8-9]. Optomechanically induced transparency (OMIT) is a property that is achieved through a dispersive coupling between vibration phonons and photons [10]. OMIT is a critical component of an integrated quantum optomechanical memory due to its ability to slow and stopping light [10-12].

Magnetically induced transparency, as well as tunable slow and fast light, have been successfully verified at room temperature [13,14]. Similarly, magnomechanically induced transparency (MMIT) resulting from the nonlinear magnon-phonon interaction in an opto-magnomechanics system [15–16] is comparable to optomechanically induced transparency. The deformation displacement of the yttrium iron garnet (YIG) bridge crystal is induced by the magnetostrictive force [17,18], and indirectly coupled to the optical cavity through radiation pressure. This hybrid system's optomechanically and magnomechanically induced transparency has huge potential applications in quantum metrology [21], quantum networks [19,20], and basic tests of quantum mechanics [22].

This work presents the Hamiltonian of the atom opto-magnomechanics system together with the associated quantum Langevin equations. The linearized quantum Langevin equations are solvable. We examine the absorption and dispersion characteristics of a weak probe field within the hybrid system. Additionally, we present the group delay τ under various magnon-phonon coupling conditions.

## 2. The model

We construct an atom opto-magnomechanical system comprising an optical cavity mode, a magnon mode, a phonon mode, and an ensemble of two-level atoms, as seen in Fig. 1. A ferromagnetic bridge is subjected to an external magnetic field on the surface of the mirror [18,23]. This generates the ferromagnetic resonance or magnon mode which is affected by the vibrations of the beam. Magnons are characterized by the collective motion of the massive spins within a ferrimagnet bridge. The magnons couple to phonons via magnetostrictive interaction. The deformation of the geometric structure of the YIG crystal is a result of the magnetization variation caused by the excitation of magnons [24-26]. This phenomenon is initiated by subjecting the crystal to a uniform bias magnetic field. An ensemble of two-level atoms is offresonantly coupled by acollective Tavis-Cummings-type interaction to the two optical fields [27]. The optical cavity is coupled to the YIG bridge's movement by radiation pressure. Frequencies and linewidths of the system as shown in Fig. 2. If the optical cavity mode c is resonant with the blue (anti-Stokes) sideband, and atomic frequencies are resonant with the red (Stokes) sideband [16].

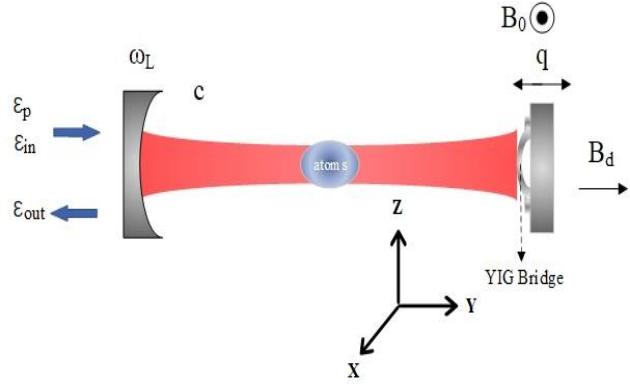

FIG. 1. The optical cavity mode (c) driven by a laser at frequency $\omega_L$ couples to an ensemble of two-level atoms (a), and a magnon mode (m) in a YIG bridge couple to a mechanical vibration mode (b) via magnetostriction. $B_0$ is the bias magnetic field (z direction), $B_d$ is the drive magnetic field (y direction).

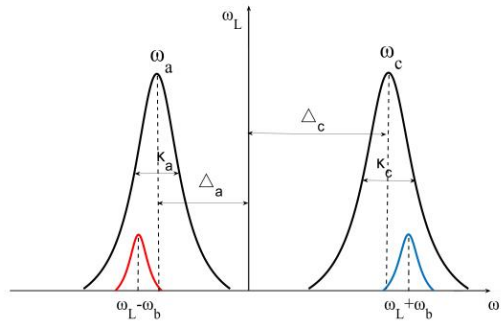

FIG.2. The frequency diagram of the system. The frequency of ensemble of two-level atoms is $\omega_a$, which is driven by a laser field at frequency $\omega_L$. And the frequency of mechanical motion is $\omega_b$, which scatters photons onto the two sidebands at $\omega_L \pm \omega_b$.

The Hamiltonian of the system under rotating-wave approximation in a frame rotating with the frequency of the drive field is given by

$$H_0/\hbar = \omega_c c^\dagger c + \omega_m m^\dagger m + \omega_b b^\dagger b$$
$$-g_c c^\dagger c(b^\dagger + b) + g_m m^\dagger m(b^\dagger + b) + H_{dri}$$
$$+(\omega_a S_Z)/2 + g_a(c_j^\dagger S_- + S_+ c_j) \tag{1}$$

where $O = c, m, b$ ($O^\dagger$) are the annihilation (creation) operators of the optical cavity mode, the magnon mode, and the phonon mode, satisfying the bosonic relation $[O, O^\dagger]=1$, $\omega_c$, $\omega_m$ and $\omega_b$ are the frequency of the optical cavity, magnon and phonon. The frequency of the magnon mode $\omega_m = \gamma H$ is determined by the bias magnetic field $H$ where $\gamma/2\pi = 28$ GHz/T denotes the gyromagnetic ratio. The fourth (fifth) term denotes the coupling between the optical cavity mode (magnon mode) and the phonon mode. $g_c$ ($g_m$) represents denotes the bare optomechanical (magnomechanical) coupling strength. The driving Hamiltonian is represented by the sixth term, $H_{dri} = i\Omega\hbar(m^\dagger e^{-i\omega_0 t} - m e^{i\omega_0 t}) + iE\hbar(c^\dagger e^{-i\omega_L t} - c e^{i\omega_L t})$ where $E = \mu\sqrt{2P_L\kappa_C/(\hbar\omega_L)}$ represents the coupling strength between the optical cavity and the laser drive field. $P_L$ ($\omega_L$) represents the power(frequency) of the laser, and $\kappa_C$ the cavity decay rate [28].

he frequency of collective spin operator of an ensemble of $N_a$ two level atoms is $\omega_a$. $S_{\pm,Z} = \sum_{j=1}^{N_a} \sigma_{\pm,Z}^{(i)}$ with $\sigma_{\pm,Z}$ being the Pauli matrices, which satisfy relations $[S_+, S_-] = S_Z$ and $[S_Z, S_\pm] = -2S_\pm$. $\sigma_+, \sigma_-$ and $\sigma_Z$ denote the spin -1/2 algebra of Pauli matrices. The atom cavity coupling constant is defined as $g_a = \mu\sqrt{\omega_c/(2\hbar\varepsilon_0 V_C)}$, $V_C$ is the optical cavity mode volume and $\mu$ is the dipole moment of the atomic transition. We consider a simplified version of the Hamiltonian which is valid in the low excitation limit. When all the atoms are initially prepared in their ground state, so that $S_Z \simeq \langle S_Z \rangle \simeq -N_a$ this condition remains largely unaffected by the interaction with the optical cavity. In this limit, the dynamics of atomic polarization can be characterized using bosonic operators. In this limit, the dynamics of atomic polarization can be described in terms of bosonic operators. In fact, $a = S_-/\sqrt{|\langle S_Z \rangle|}$ is defined as the atomic annihilation operator, and it satisfies the usual bosonic commutation relation $[a, a^\dagger] = 1$ [29]. We obtain the fully simplified Hamiltonian, given by

$$H_1/\hbar = \omega_c c^\dagger c + \omega_m m^\dagger m + \omega_b b^\dagger b$$
$$-g_c c^\dagger c(b^\dagger + b) + g_m m^\dagger m(b^\dagger + b) + H_{dri}$$
$$+\omega_a a^\dagger a + g_a(c^\dagger a + a^\dagger c) \tag{2}$$

We establish a set of quanta Langevin equations of the system:
$$\dot{a} = -(i\Delta_a + \kappa_a)a - ig_a c + \sqrt{2\kappa_a}a^{in}$$
$$\dot{c} = -(i\Delta_c + \kappa_c)c - ig_a a + g_c(b^\dagger + b) + \varepsilon_d e^{i\delta t} + \sqrt{2\kappa_c}c^{in}$$
$$\dot{m} = -(i\Delta_m + \kappa_m)m - ig_m m(b^\dagger + b) + \Omega + \sqrt{2\kappa_m}m^{in}$$
$$\dot{b} = -i\omega_b b - ig_m m^\dagger m - \kappa_b b + ig_c c^\dagger c + \sqrt{2\kappa_b}b^{in} \tag{3}$$

Where $\Delta_{a(c)} = \omega_{a(c)} - \omega_L$ and $\Delta_m = \omega_m - \omega_0$, $\kappa_c$, $\kappa_m$, and $\kappa_b$ are the dissipation rates of the cavity, magnon, mechanical modes, $\kappa_a$ is the decay rate of the two-level atoms excited level. Respectively, and $o^{in}$ (o=a, m, c, b) are zero-mean input noise operators [30]. The linear expansion denotes the operator as the sum of quantum fluctuation and steady-state value $o = o_s + o_+ e^{i\delta t} + o_- e^{i\delta t}$. the steady-state solutions of the dynamic operator are obtained as

$$(i\lambda - \kappa_a)a_+ - ig_a c_+ = 0$$
$$(i\lambda - \kappa_c)c_+ - ig_a a_+ + ig_c b_+ + \varepsilon_d = 0$$
$$(i\lambda - \kappa_m)m_+ - ig_m b_+ = 0$$
$$i\lambda b_+ - ig_m m_+ - \kappa_b b_+ + ig_c c_+ = 0 \tag{4}$$

we take $\lambda = \delta - \omega_b$ and $\Delta_a = \Delta_c = \Delta_m = \omega_b$. Therefore, we can get the output field response as

$$c_+ = \frac{\varepsilon_p}{(i\lambda-\kappa_c)+\frac{g_a^2}{(i\lambda-\kappa_a)}+\frac{g_c^2}{\left(i\lambda-\kappa_b+\frac{g_m^2}{(i\lambda-\kappa_m)}\right)}} \quad (5)$$

Satisfying the input-output relation: $\varepsilon_{out} = \varepsilon_{in} - 2\kappa_c c_+$ [31]. The amplitude of the output field oscillating can be written as

$$\varepsilon_{out} = \frac{2\kappa_c c_+}{\varepsilon_p} \quad (6)$$

The real (imaginary) part of $\varepsilon_{out}$ shows the absorption (dispersion) of the detected field.

### 3. Analysis and discussion

Mechanical modes in the hybrid system need to be cooled to the ground state, because of the high heat density of mechanical modes at low temperatures. In a narrow band surrounding resonance, the system can be converted from absorptive to transmittive by employing an ensemble of two-level atoms. This effect results in an exceptionally steep dispersion of the probe photons that are transmitted [11]. We used experimentally feasible parameter values: $\omega_m/2\pi = 10$ GHz, $\omega_b/2\pi = 40$ MHz, $\gamma_b/2\pi = 10^2$ Hz, $\kappa_a/2\pi = \kappa_m/2\pi = 1$ MHz, $\kappa_c/2\pi = 2$ MHz, $\lambda_L = 1064$ nm (optical wavelength), $\Delta_m = \omega_b$. [27]

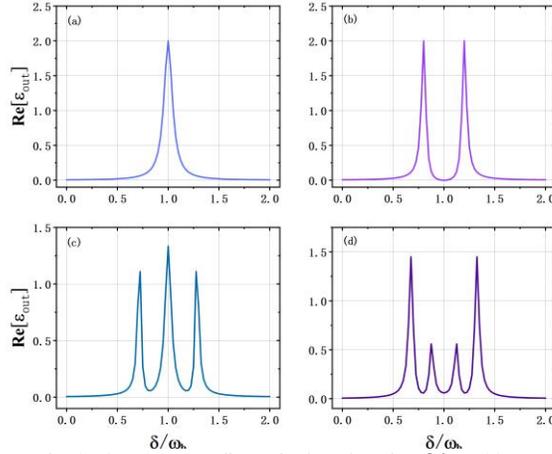

FIG.3. Polt of absorption Re[$\varepsilon_{out}$] versus dimensionless detuning $\delta/\omega_b$. (a) $g_a = g_m = g_c = 0$, (b) $g_a = g_m = 0$, $g_c/2\pi = 8\ MHz$, (c) $g_a = 0$, $g_c/2\pi = g_m/2\pi = 8\ MHz$, (b) $g_m/2\pi = g_c/2\pi = g_a/2\pi = 8\ MHz$.

Fig. 3(a) represents the absorption spectrum of a single optical cavity. In the absence of magnetostrictive contacts and a collection of two-level atoms, a standard Lorentzian peak bifurcates into a twofold window as a result of the opto-mechanical coupling illustrated in Fig. 3(b). As shownin figure 3(c), we observe three transparency windows in the absorption as we switch on the magnon-phonon coupling and photon-phonon coupling. Due to the coherent coupling between the upper and lower energy states of the two energy level atoms in the presence of detected and signal light, a width magnomechanically induced transparency window splits into two narrow windows in the system in Fig. 3(d).

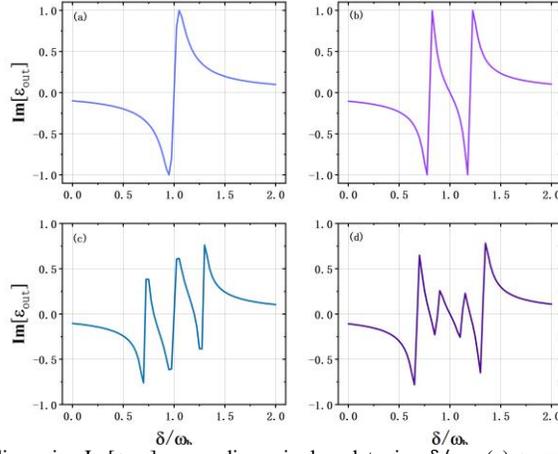

FIG.4. Polt of dispersion Im[$\varepsilon_{out}$] versus dimensionless detuning $\delta/\omega_b$. (a) $g_a = g_m = g_c = 0$, (b) $g_a = g_m = 0$, $g_c/2\pi = 8\ MHz$, (c) $g_a = 0$, $g_c/2\pi = g_m/2\pi = 8\ MHz$, (b) $g_m/2\pi = g_c/2\pi = g_a/2\pi = 8\ MHz$.

We plot the dispersion spectrum of the output field against the normalized frequency of the field. The dispersion spectrum of single optical cavity is shown in Fig. 4(a). When only optomechanical interactions exist in the system, we observed two peaks in Fig. 4(b). The dispersion spectra for the case of $g_c \neq 0$, $g_m \neq 0$ and $g_a = 0$ is plotted in the Fig. 4(c). Fig. 4(d) represents the output spectrum in the presence of an ensemble of two-level atoms.

From Eq. (6), the rescaled transmitted field corresponding to the probe field can be expressed as

$$t_p = \frac{\epsilon_p - 2\kappa_c c_+}{\epsilon_p} \quad (7)$$

The transmission spectrum of the probe field is plotted versus the scaled detuning $\delta/\omega_b$ in Figure 5. Figs. 5(a) and 5(b) display the transmission curve without the atomic ensemble. Without the atomic ensemble, the curve of transmission is shown in Fig. 5(a) and 5(b). This is due to the fact that the coupling in Fig. 5(b) is weaker than that of $g_m$. The transmission curve is displayed in Figs. 5(c) and 5(d) when an atomic ensemble is inserted into the cavity. By contrasting Figs. 5(a) and 5(b), as well as Figs. 5(c) and 5(d), it can be seen that the transmission spectrum widens in proportion to the rise in magnon-phonon coupling.

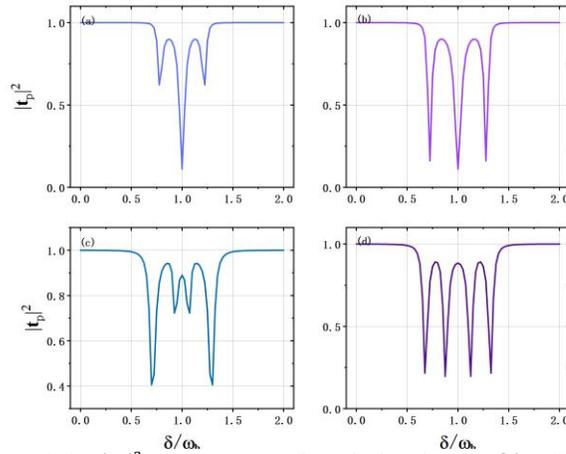

FIG.5. The transmission $|t_p|^2$ spectrum versus dimensionless detuning $\delta/\omega_b$. We take $g_a = 0$ in (a) and (b), $g_m/2\pi = 4\ MHz$ in (a) and (c). Other parameters follow the above mentioned.

he phase $\phi$ of the transmitted probe field $t_p$ is defined by $\phi = Arg[t_p]$. The group delay $\tau$ of the output field is given by

$$\tau = \frac{\partial \phi}{\partial \omega_p} = Im[\frac{1}{\varepsilon_{out}}\frac{\partial \varepsilon_{out}}{\delta \delta}] \qquad (8)$$

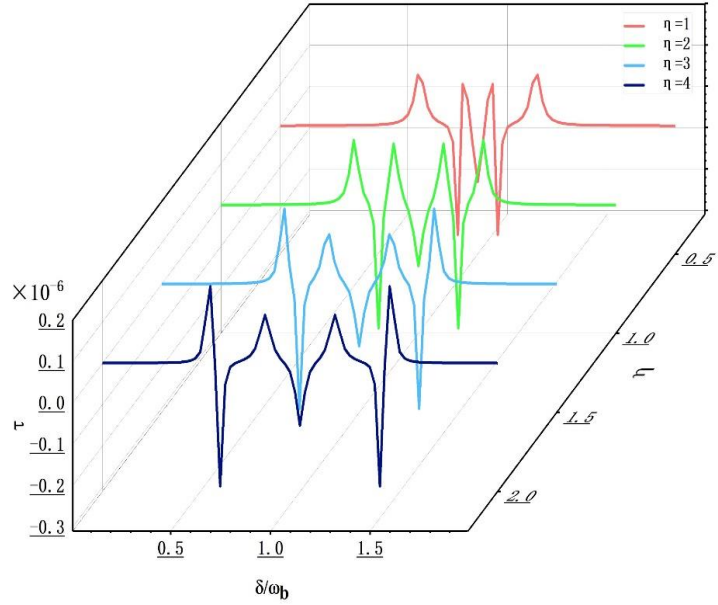

FIG.6. 3D Polt of the group delay τ versus dimensionless detuning $\delta/\omega_b$ and η. We take $\eta = g_{m/}g_c$ and $g_a/2\pi = 8$MHz.

Figure 6 shows that the group delay τ can be tuned by the magnomechanical coupling values. This group delay τ can be increased by raising the $G_{mb}$, and the time delay for slow light is increased up to 0.2 ms. We observed longer group time delays in the opto-magnomechanics system. We demonstrate tunable optical delays in nanoscale magnomechanical crystals. When the frequency of the light is close to resonance $\delta/\omega_b = 1$, the group delay τ becomes negative in this regime.

## 4. Conclusion

In an atom opto-magnomechanics system, we have examined the transmission and absorption spectrum of a weak probe field under a strong field. In this system, we examine the output field's absorption and dispersion spectra under various coupling scenarios. A nonlinear phonon-magnon interaction was shown to be responsible for the magnomechanically induced transparency (MMIT) that we observed. Additionally, optomechanically induced transparency (OMIT) is a result of the photon-phonon interaction. When an ensemble of two-level atoms is placed in the system, the absorption propertie is flipped. We investigate the transmission of the output signal and show the effect of the two-level atoms and magnon-phonon coupling on the transmission spectrum. It is demonstrated that varying phonon-magnon coupling strength values correspond to varied group delay τ. Our solutions have great potential in the field of quantum precision measurements.

**Funding.** Project supported by the National Natural Science Foundation of China (Grant No.62071376, 62005211 、52175531) 、Natural Science Foundation of Chongqing, China （NO.CSTB2024NSCQ-MSX0746） 、Sustainedly Supported Foundation by National Key Laboratory of Science and Technology on Space Microwave under Grant HTKJ2024KL504002, State

Key Laboratory of Quantum Optics and Quantum Optics Devices（No：KF202408）, the National Key Research and Development Program of China (Grant No. 2021YFC2203601).

**Disclosures.** The authors declare no conflicts of interest.

**Data availability.** The data that support the findings of this study are available within the article.